\documentclass[a4paper,12pt]{article}
\pdfoutput=1
\usepackage{amsmath}
\usepackage{amsfonts}
\usepackage{amssymb}
\usepackage{graphicx, rotating}
\usepackage{epsfig}
\usepackage{latexsym}
\usepackage{xcolor}
\usepackage{bm}

\usepackage[english]{babel}
\usepackage[colorlinks,allcolors=blue]{hyperref}
\usepackage[numbers,sort&compress]{natbib}

\newcommand{\Slash}[1]{{\ooalign{\hfil#1\hfil\crcr\raise.167ex\hbox{/}}}}

\newcommand{\beq}{\begin{equation}}  \newcommand{\eeq}{\end{equation}}
\newcommand{\bef}{\begin{figure}}  \newcommand{\eef}{\end{figure}}
\newcommand{\bec}{\begin{center}}  \newcommand{\eec}{\end{center}}
  
\newcommand{\laq}[1]{\label{eq:#1}}  

\newcommand{\Eq}[1]{Eq.~(\ref{eq:#1})}

\newcommand{\eq}[1]{(\ref{eq:#1})}

\def\({\left(}
\def\){\right)}

\newcommand{\MEV}{ {\rm \, MeV} }
\newcommand{\GEV}{ {\rm \, GeV} }

\def\a{\alpha}

\def\f{\phi}
\def\g{\gamma}

\def\m{\mu}

\def\s{\sigma}

\def\D{\Delta}

\def\F{\Phi}

\def\*{\dagger}

\begin{document}
\renewcommand\bibname{\Large References}

\begin{center}

\hfill   TU-1259 \\

\vspace{1.5cm}

{\Large\bf   Isotropic cosmic birefringence from string axion domain walls without cosmic strings,  and DESI results}\\
\vspace{1.5cm}

{\bf  
Junseok Lee$^{1}$, 
Kai Murai$^{1}$,
Fuminobu Takahashi$^{1}$, 
Wen Yin$^{2}$}

\vspace{12pt}
\vspace{0.5cm}
{\em 
$^{1}$Department of Physics, Tohoku University,  
Sendai, Miyagi 980-8578, Japan \\
$^{2}$Department of Physics, Tokyo Metropolitan University, Tokyo 192-0397, Japan
\vspace{5pt}}
\vspace{0.5cm}
\abstract{
Recently, results from the Atacama Cosmology Telescope (ACT) DR6 have shown a preference for isotropic cosmic birefringence, consistent with previous analyses based on Planck and WMAP data.
Separately, the Dark Energy Spectroscopic Instrument (DESI) DR2 results suggest that dark energy evolves over cosmic history, pointing to new physics in the late-time universe.
In this paper, we propose that domain walls associated with the string axion can naturally explain the isotropic cosmic birefringence, focusing on the case in which the axion starts near a hilltop.
Interestingly, to avoid the domain wall problem, these walls must form well after recombination.
The predicted rotation angle, $\beta \approx 0.21\,c_\gamma$ degrees (with anomaly coefficient $c_\gamma \approx 1$), is in excellent agreement with observations.
This scenario can be further tested by probing anisotropic birefringence of photons emitted long after recombination, as well as gravitational waves.
Moreover, starting the axion oscillation from a hilltop naturally enhances its abundance via anharmonic effects, thus contributing to the dark energy component.
We discuss how this hilltop axion scenario may connect with the DESI results.
}

\end{center}
\clearpage

\setcounter{page}{1}
\setcounter{footnote}{0}

\section{Introduction}
Light axions may be ubiquitous in nature. In string or M-theory, a multitude of axions commonly appears, which we denote collectively by $\phi$. Their cosmological and phenomenological implications have been thoroughly investigated within the frameworks of the axiverse and axion landscape scenarios \cite{Witten:1984dg, Svrcek:2006yi,Conlon:2006tq,Arvanitaki:2009fg,Acharya:2010zx, Higaki:2011me, Cicoli:2012sz,Demirtas:2018akl}. The axion exhibits a discrete shift symmetry, 
\begin{eqnarray}
\laq{shift}
\phi  \rightarrow \phi + 2  \pi f_\f ~,
\end{eqnarray}
where $f_\phi$ is the axion decay constant.
Unless one considers large-volume compactification~\cite{Balasubramanian:2005zx}, the string axion typically possesses a decay constant of
\beq\laq{string}
    f_\phi
    = 10^{15}\,\text{--}\,10^{17}\GEV.
\eeq
The axion potential is generally generated via non-perturbative effects, and its mass can be exponentially suppressed. A distinctive prediction of the theories involving axions is the existence of degenerate vacua connected by the discrete shift symmetry \eq{shift}; if these degenerate vacua are realized in the universe, they will be separated by domain walls.

The axion may also couple to photons through an anomaly
\begin{align}
\label{eq:int}
{\cal L} & = c_\g \frac{\a}{4 \pi} \frac{\phi}{f_\f} F_{\mu \nu} \tilde F^{\mu \nu} 
\equiv \frac{1}{4} g_{\phi\g\g} \phi F_{\mu \nu} \tilde F^{\mu \nu},
\end{align}
where $c_\g$ is an anomaly coefficient,%
\footnote{%
Note that the definition of $c_\gamma$ differs from the 
definition of the electromagnetic anomaly coefficient $E$ for the QCD axion adopted in some literature
by a factor of $2$.
Our definition of $c_\gamma$ follows the normalization of the U(1) anomaly used in, e.g., Ref.~\cite{DiLuzio:2020wdo}.
}
$\a$ is the fine structure constant, and $F_{\mu \nu}$ and $\tilde{F}_{\mu \nu}$ denote the field strength and its dual, respectively. In the literature, an axion with such a photon coupling is often referred to as an axion-like particle (ALP). For comprehensive reviews on axions and related topics, see Refs.~\cite{Jaeckel:2010ni,Ringwald:2012hr,Arias:2012az,Graham:2015ouw,Marsh:2015xka, Irastorza:2018dyq, DiLuzio:2020wdo}.
The axion photon coupling is constrained as $g_{\phi \gamma \gamma} < 6.3 \times 10^{-13}\,\mathrm{GeV}^{-1}$ or equivalently 
$f_\phi > 3.7 \times 10^9 c_\gamma$\,GeV
for axions lighter than $10^{-12}$\,eV~\cite{Reynes:2021bpe}.

Recently, a hint of isotropic cosmic birefringence (CB) in the cosmic microwave background (CMB) polarization was reported, with a measured rotation angle~\cite{Louis:2025tst}
\beq
\laq{measure}
\beta= 0.20\pm 0.08 {\rm~ deg },
\eeq
based on an analysis of the ACT DR6 polarization data. 
This is consistent with previous analyses of the Planck and WMAP data $\beta=0.34\pm 0.09~{\rm deg}$~\cite{Eskilt:2022cff} (see also Refs.~\cite{Minami:2020odp,Diego-Palazuelos:2022dsq,Eskilt:2022wav,Cosmoglobe:2023pgf}).
Assuming independence, we combine the results and obtain 
\beq
\beta_{\rm combined}=0.26\pm 0.06~{\rm deg},
\eeq achieving an excess greater than $4\s$.

One plausible mechanism to generate this CB is to introduce an ALP that varies temporally and/or spatially~\cite{Carroll:1998zi,Finelli:2008jv,Arvanitaki:2009fg,Panda:2010uq,Fedderke:2019ajk,Fujita:2020aqt,Fujita:2020ecn,Takahashi:2020tqv,Mehta:2021pwf,Nakagawa:2021nme,Choi:2021aze,Obata:2021nql,Yin:2021kmx,Kitajima:2022jzz, Gasparotto:2022uqo,Lin:2022niw,Murai:2022zur,Gonzalez:2022mcx, Galaverni:2023zhv,Eskilt:2023nxm,Yin:2023srb,Naokawa:2023upt,Murai:2023xjn,Gasparotto:2023psh,Gendler:2023kjt,Greco:2024oie,Lee:2024oaz,Naokawa:2024xhn,Lee:2024xjb, Murai:2024yul,Zhang:2024dmi,Kochappan:2024jyf,Blasi:2024xvj,Yin:2024pri} (see also Ref.~\cite{Komatsu:2022nvu} for a review). 
When a photon propagates through a slowly varying ALP background, its polarization angle $\Phi$ evolves as~\cite{Carroll:1989vb,Carroll:1991zs,Harari:1992ea}
\beq
\laq{delF}
\dot{\F}\approx \frac{ c_\g \a}{2\pi}\frac{{ \hat{r}^\mu \partial_\m \f}}{f_\f},
\eeq
where the dot denotes a derivative with respect to time along the photon trajectory, and $\hat{r}^\mu$ represents the normalized photon four-momentum; for example, $\hat{r}=(1,0,0,1)$ when the photon travels in the positive $z$ direction. 
By integrating this equation along the line of sight from the last scattering surface (LSS) to the observer, one obtains a net rotation of the polarization plane
\beq
\D\F(\Omega) = 
  0.42 {\rm~ deg } \times c_\g \( \frac{\f_{\rm today}-\f_{\rm LSS}(\Omega) }{2\pi f_\phi}\),
\eeq
where $\f_{\rm today}$ and $\f_{\rm LSS}(\Omega)$ denote the axion field values at the solar system today and at the LSS, respectively, and $\Omega$ specifies the angular direction in polar coordinates $(\theta,\varphi)$. The isotropic CB angle is then given by
\beq
\beta= \frac{1}{4\pi} \int{ d\Omega \,{\D \F}(\Omega)}.
\eeq
If the change in the axion field value exactly corresponds to half of the discrete shift symmetry transformation \eq{shift}, i.e., if $\f_{\rm today}-\f_{\rm LSS}(\Omega) = \pi f_\phi$, then the rotation angle becomes 
\begin{equation}
    \beta \;\simeq\; 0.21 c_\gamma \,{\rm deg},
    \laq{pred}
\end{equation}
which is intriguingly close to the observed value \eq{measure} for $c_\gamma = 1$.
This coincidence motivates us to focus on the possibility that the axion initially resides near the hilltop of the potential, i.e., midway between adjacent vacua related by the periodicity \eq{shift}.

If the axion initially resides near the hilltop of the potential, it begins to evolve after the Hubble parameter becomes comparable to its mass.%
\footnote{The anharmonic effect, which modifies the axion dynamics near the hilltop, will be discussed later.}
At this stage, two possibilities arise. One possibility is that fluctuations of the axion do not overcome the potential maximum, so the field remains nearly homogeneous and begins to oscillate around one of the potential minima. In this case, the isotropic CB is predicted by \Eq{pred}. The other possibility is that axion fluctuations cross the potential maximum, leading to the population of both vacua. As a result, domain walls form between these vacua.

In particular, it was proposed by\footnote{Two of the present authors (FT and WY).} Ref.~\cite{Takahashi:2020tqv} that the reported value of the rotation angle can be explained by ALP domain walls in the scaling solution.
In this scenario, the region around Earth lies within a single domain, while the domain wall network at recombination contains $10^{3\text{--}4}$ domains across the LSS. As a result, the axion field value on the LSS effectively averages to the midpoint between two adjacent vacua.
Depending on which vacuum the axion occupies in a given domain, the rotation angle $\D \F(\Omega)$ takes one of two possible values—either $0$ or $0.42\,c_\gamma$ degrees.
Consequently, each domain contributes one bit of information to the CB,
and this phenomenon has been termed kilobyte cosmic birefringence (KBCB).
Remarkably, spatial averaging over many domains yields an isotropic CB prediction that is nearly identical to that in the case where the axion starts uniformly near the hilltop and evolves coherently, as given by \Eq{pred}.

Since the two vacua are physically equivalent and related by the discrete shift symmetry \Eq{shift}, one cannot introduce a potential bias to lift the degeneracy, and thus the domain wall problem cannot be avoided in this way.
To avoid this problem when formation occurs before recombination, the KBCB scenario requires the axion to have a decay constant $f_\phi \ll 10^{15}\GEV$, which lies outside the typical range for string axions and therefore necessitates an explanation for the origin of the ALP.
Furthermore, the formation of domain walls requires the axion field to acquire fluctuations around the potential maximum.
Such fluctuations can naturally arise for string axions during inflation and typically result in a population bias.\footnote{Note that domain wall networks seeded by inflationary fluctuations remain stable under the population bias, in contrast to those generated by thermal or white-noise fluctuations~\cite{Gonzalez:2022mcx}.}
This bias distorts the otherwise equal distribution of the two vacua on the LSS, shifting the predicted value of $\beta$ and rendering the simple averaging approximation inaccurate.

In this paper, we focus on a string axion with a large decay constant \eq{string}, initially located near the hilltop of its potential. This setup is motivated by the striking agreement between the observed rotation angle \eq{measure} and the theoretical prediction \eq{pred} for $c_\gamma = 1$.
In such a scenario, we can consider two possibilities as mentioned above: the axion oscillates homogeneously or forms domain walls.
Since the former case has already been discussed and leads to the simple prediction given in \eq{pred}, we concentrate in this paper on the latter possibility with domain walls.
In this case, their formation must occur after recombination in order to avoid the domain wall problem.
This requirement significantly suppresses the uncertainty in the predicted value of $\beta$ that would otherwise arise from population bias, as in the KBCB scenario.
We therefore concentrate on post-recombination domain wall formation, which presents distinctive features for explaining the observed CB with a string axion.

On the other hand, the Dark Energy Spectroscopic Instrument (DESI) DR2 baryonic acoustic oscillation measurements~\cite{DESI:2025zgx,DESI:2025zpo}, based on a larger and higher-quality dataset than DR1~\cite{DESI:2024mwx}, improve the statistical precision by about $40\%$ as stated in Sec.~VI of Ref.~\cite{DESI:2025zgx}.
Combined with CMB and supernova data~\cite{Rubin:2023ovl,DES:2024jxu}, these results provide stronger hints—at the $2.8\sigma$ to $4.2\sigma$ level—for a time-dependent dark energy component, beyond a simple cosmological constant.

Various explanations for this possibility have been proposed~\cite{Tada:2024znt,Yin:2024hba,Cortes:2024lgw,Bhattacharya:2024hep,Mukherjee:2024ryz,Notari:2024rti,Jia:2024wix,Hernandez-Almada:2024ost,Bhattacharya:2024kxp,Berbig:2024aee,Wolf:2025jlc,Chakraborty:2025syu,Borghetto:2025jrk}. Among them, models of time-varying dark energy motivated by CB have been studied in Refs.~\cite{Tada:2024znt,Yin:2024hba,Berbig:2024aee} (see also Ref.~\cite{Choi:2021aze} before the DESI data release and earlier studies~\cite{Linder:2002et,dePutter:2008wt}).
Such time-dependent dark energy may be connected to the cosmological constant problem~\cite{Yin:2021uus,Yin:2024hba}, and indeed, it arises naturally in models that address it~\cite{Yin:2021uus}. In this context, a hilltop axion model with matter effects to also address the $S_8$ tension has been proposed~\cite{Khoury:2025txd} (see also~\cite{Gonzalez:2022mcx} for domain wall formation and~\cite{Daido:2017wwb,Co:2018mho,Takahashi:2019pqf,Nakagawa:2020eeg,Narita:2023naj,Co:2024bme} for general hilltop axion models).

In our scenario, the axion starts near the potential maximum, where anharmonic effects enhance its abundance. As a result, even with a decay constant well below the Planck scale, the resulting oscillation energy can make a non-negligible contribution to the dark energy component. We investigate the implications of this scenario for the DESI results, focusing on the dark energy contribution from the hilltop axion.

\section{String axion domain walls and isotropic cosmic birefringence}
\label{sec:2}

In this section, we investigate a scenario in which an axion initially located near the hilltop of its potential leads to the formation of domain walls without strings, and explore the resulting implications for CB.
For our purposes, it is sufficient to consider the following potential:
\begin{eqnarray}
\label{pot}
V(\phi) &=& \Lambda^4 \left[ 1 + \cos\left(\frac{\phi}{f_\phi}\right) \right] \nonumber \\
&\simeq& 2\Lambda^4 - \frac{1}{2} m_{\phi}^2 \phi^2 + \frac{1}{4} \left(\frac{\Lambda}{f_\phi}\right)^4 \phi^4 + \cdots,
\end{eqnarray}
where $\Lambda$ is a dynamical scale, $m_{\phi} \equiv \Lambda^2/f_\phi$ is the mass of the axion.
The second line represents an expansion around the origin. 
This potential is invariant under the discrete shift symmetry transformation \eq{shift}, and consequently admits degenerate vacua. One of the maxima is located at the origin, and there are degenerate vacua at $\phi = \pm \pi f_\phi, \pm 3 \pi f_\phi, \cdots$. In the following we consider a case in which $\phi$ is initially around the origin so that the dynamical range of $\phi$ is limited to $- 2 \pi f_\phi < \phi < 2 \pi f_\phi$.

Let us assume that during inflation the axion is nearly massless, i.e., $H_{\rm inf} \gg m_\phi$, where $H_{\rm inf}$ is the Hubble parameter during inflation. In this case, the axion acquires quantum fluctuations with amplitude 
\[
\delta\phi = \frac{H_{\rm inf}}{2\pi}
\]
around a zero mode $\phi_0$ at the time of horizon exit, and these fluctuations become classical afterward. Here, the zero mode refers to the axion field value averaged over the observable universe. 
The superhorizon fluctuations accumulate in a random-walk fashion, and the variance of the Gaussian probability distribution of the axion field, $\sigma_\phi^2$, grows as 
\[
\sigma_\phi^2(N_e) = N_e\left(\frac{H_{\rm inf}}{2\pi}\right)^2,
\]
where $N_e$ is the number of e-folds. After inflation, the axion begins to oscillate when the Hubble parameter drops to approximately 
\begin{align}
\label{osc}
    H \sim H_{\rm osc} = m_\phi.
\end{align}
%
Note that the onset of oscillations could be delayed due to the flat shape of the potential when $|\phi| \ll f_\phi$, namely around the hilltop.
This anharmonic effect delays the domain wall formation as discussed below.
Suppose that $\phi_0$ is sufficiently close to the hilltop at the origin such that
\beq 
\laq{cond} 
|\phi_0| \lesssim \sigma_\phi(N_{\rm osc}), 
\eeq
where $N_{\rm osc}$ denotes the number of e-folds before the end of inflation at which the mode with wavenumber $k/a = H_{\rm osc}$ exits the horizon.
This condition is essential for the scenario to work and will be justified later.
Since the probability distribution of the axion spans both sides of the potential maximum, domain walls are formed when the axion begins to oscillate.
Although inflationary fluctuations naturally induce a population bias, the large-scale correlation of superhorizon modes renders the domain wall network long-lived~\cite{Gonzalez:2022mcx}.

\begin{figure}[!t]
\begin{center}  
   \includegraphics[width=145mm]{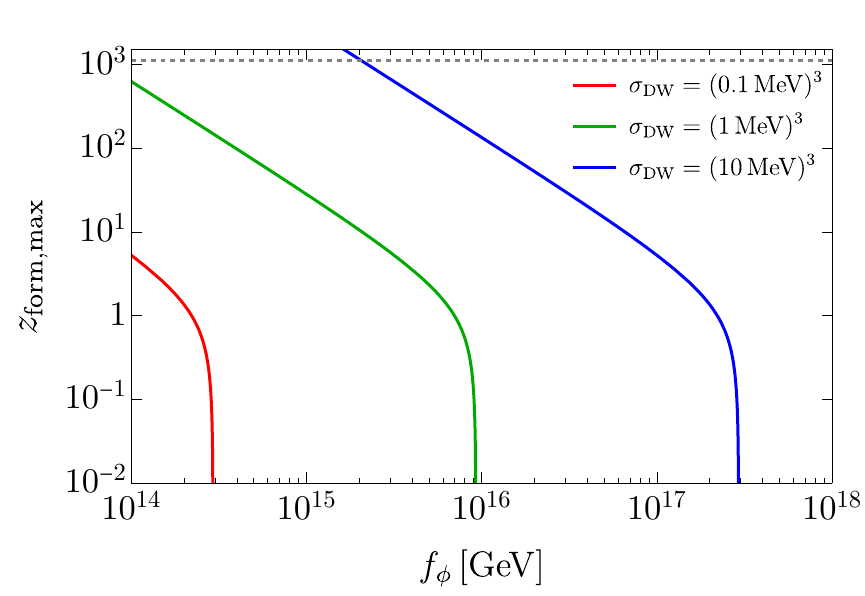}
\end{center}
\caption{
    Redshift of domain wall formation estimated using Eq.~\eqref{osc} for $\sigma_{\rm DW} = 8 m_\phi f_\phi^2 = (0.1\,\mathrm{MeV})^3$, $(1\,\mathrm{MeV})^3$, and $(10\,\mathrm{MeV})^3$ from left to right. 
    The horizontal dashed line denotes the recombination epoch, $z \sim 1100$.
    Given the possible delay due to the anharmonic effect (see Fig.~\ref{fig:2}), this estimate is regarded as the maximal redshift of the domain wall formation.
}
\label{fig:1}
\end{figure}
\begin{figure}[!t]
\begin{center}  
   \includegraphics[width=81mm]{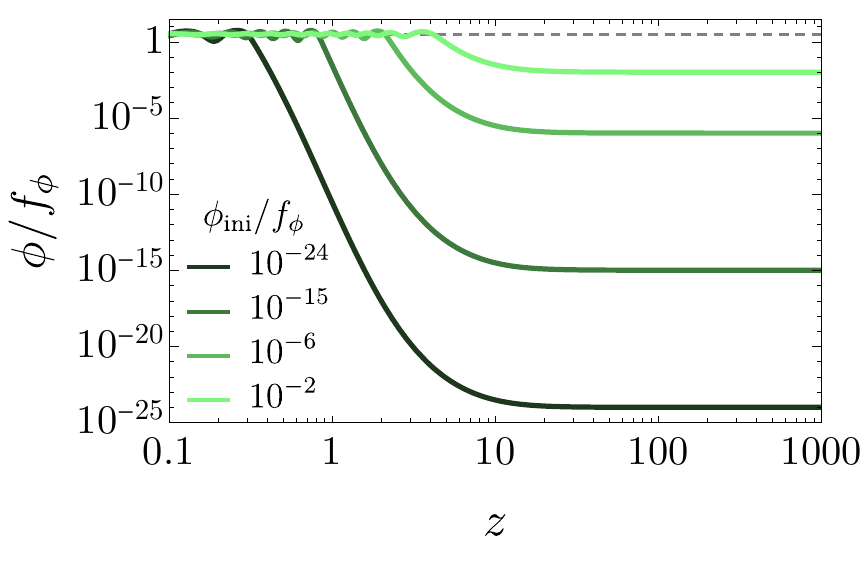}
   \includegraphics[width=81mm]{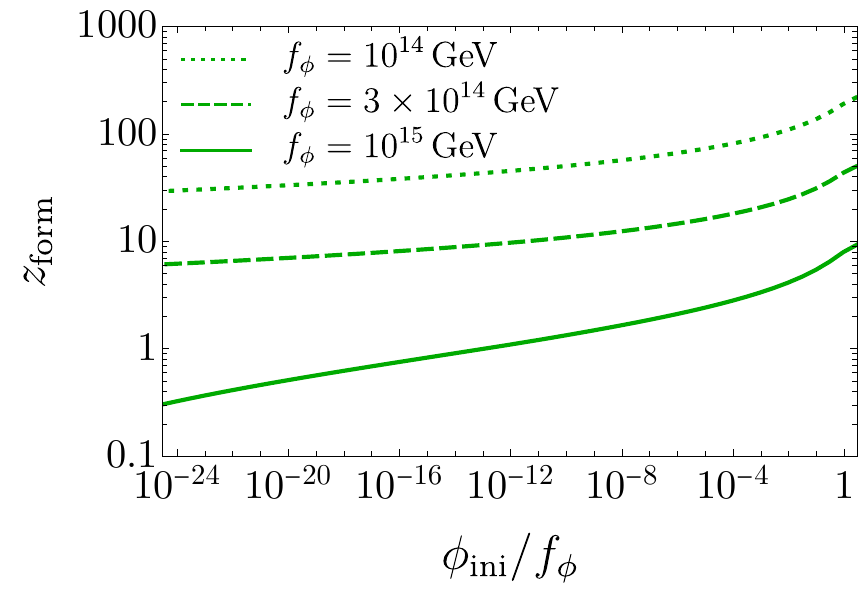}
\end{center}
\caption{%
    Delay of the domain wall formation with hilltop initial conditions for $\sigma_{\rm DW} = (1\,\MEV)^3$, corresponding to the green line in Fig.~\ref{fig:1}.
    \textit{Left panel}:
    Evolution of the axion field $\phi$ for $f_\phi = 10^{15}\,\GEV$ with different initial conditions, $\phi_{\rm ini}/f_\phi = 10^{-24}$, $10^{-15}$, $10^{-6}$, $10^{-2}$ from bottom to top.  
    These represent typical displacements from the potential maximum, expected to be of order $H_{\rm inf}/f_\phi$.  
    The horizontal dashed line denotes $\phi/f_\phi = \pi$. 
    \textit{Right panel}:
    Redshift at the domain wall formation, $z_\mathrm{form}$, defined as the time when $\phi$ first crosses $\pi f_\phi$, as a function of $\phi_{\rm ini}/f_\phi$.
    We adopt the decay constants of $f_\phi = 10^{14}$\,GeV (dotted), $3 \times 10^{14}$\,GeV (dashed), and $10^{15}$\,GeV (solid).
    The solid line corresponds to the left panel.
    }
\label{fig:2}
\end{figure}

Domain walls form shortly after the axion begins to oscillate, at the time when condition~(\ref{osc}) is satisfied.
The corresponding redshift at the onset of oscillations, estimated using this condition, is shown in Fig.~\ref{fig:1}.
Here we use the fact that the domain wall tension is given by $\sigma_{\rm DW} \simeq 8 m_\f f_\f^2$ for the potential (\ref{pot}) and fix the tension by $\sigma_{\rm DW} = (0.1\,\mathrm{MeV})^3$, $(1\,\mathrm{MeV})^3$, and $(10\,\mathrm{MeV})^3$, from left to right.
Once we fix $\sigma_\mathrm{DW}$, we can relate $f_\phi$ to $m_\phi$, and then to the time of the domain wall formation using Eq.~\eqref{osc}.
Note that this estimate is valid when the field fluctuations are not much smaller than $f_\phi$ or when the zero mode is not very close to the hilltop.
If the fluctuations are small and the zero mode is in the vicinity of the hilltop, the anharmonic effect becomes significant and delays the onset of oscillations.
Fig.~\ref{fig:2} illustrates how the anharmonic effect causes this delay. As shown in Fig.~\ref{fig:2}, the initial field value of the axion must be extremely close to the potential maximum in order to significantly delay the onset of domain wall formation.
In this sense, the redshift shown in Fig.~\ref{fig:1} should be interpreted as the maximal redshift of the domain wall formation.

The cosmological domain wall problem imposes the following upper bound on the tension~\cite{Zeldovich:1974uw,Vilenkin:1984ib}:
\beq \label{tension} \sigma_{\rm DW} \simeq 8 m_\f f_\f^2 \lesssim (1 \MEV)^3. \eeq
This bound arises from the constraint on CMB temperature anisotropies induced by the gravitational potential of domain walls.
Given this constraint, the decay constant is limited to $f_\phi \lesssim 10^{16}\,\mathrm{GeV}$.
In particular, if we focus on the range of $f_\phi$ typical for a string axion, as given in \eq{string}, we find that domain wall formation must occur well after recombination—specifically at redshift $z_{\rm form} \lesssim 100$.

Since $H_{\rm inf} \lesssim 10^{13}\,\mathrm{GeV}$ from current CMB constraints~\cite{BICEP:2021xfz,Tristram:2021tvh,Campeti:2022vom,Galloni:2022mok}, and taking into account \Eq{string}, we find that at recombination the condition \eq{cond} reads
\beq \frac{|\bar \f_{\rm LSS}|}{f_\f} < \frac{\s_\f}{f_\f} \simeq \sqrt{N_{e}} \frac{H_{\rm inf}}{2\pi f_{\phi}} \ll 1, \eeq
where $\bar \f_{\rm LSS}$ denotes the angular average of $\f_{\rm LSS}(\Omega)$ over the LSS, and we have used the fact that $N_e \simeq 50 \sim 60$.
Since the observer resides in one of the two vacua after domain wall formation, the isotropic CB is given by
\beq 
\boxed{ \beta = 0.42\,\mathrm{deg} \times c_\g \left( \frac{\f_{\rm today}-\f_{\rm LSS}}{2\pi f_\phi} \right) \approx 0.21\,\mathrm{deg} \times c_\g \left( \frac{\f_{\rm today}}{\pi f_\phi} \right)}, 
\label{eq: beta formula}
\eeq
where the sign in the second expression is chosen to be positive to match the observed sign of the rotation angle, namely, we take $c_\gamma \phi_{\rm today}>0$. 

Note that the origin of the isotropic CB in this scenario differs from that in the original KBCB mechanism.  
There, domain wall formation has already occurred by the time of last scattering, and both adjacent vacua are populated on the LSS.  
The average axion field is given by $\bar{\phi}_{\rm LSS} = (\phi_{\rm min,1} + \phi_{\rm min,2})/2$, where $\phi_{\rm min,1}$ and $\phi_{\rm min,2}$ denote the field values of the adjacent minima, one of which corresponds to our local vacuum.  
The isotropic CB then arises from averaging over these domains, while anisotropic CB is also generated due to spatial variations in the field.
In this case, if we take into account the population bias between the two vacua, the predicted rotation angle may carry uncertainty depending on the significance of the population bias.
In contrast, in the present scenario, the axion remains near the hilltop until well after recombination, and the field on the LSS is nearly homogeneous.
Consequently, the CB from CMB photons, which is induced by the difference in $\phi_\mathrm{LSS}$ and $\phi_\mathrm{today}$, is purely isotropic.
Then, the isotropic CB angle is determined by Eq.~\eqref{eq: beta formula} and is free from ambiguity related to population bias.
Remarkably, despite the differences in the underlying dynamics, the predicted isotropic rotation angle is the same in both scenarios.

To satisfy the condition in \Eq{cond}, several possibilities can be considered.
One is the presence of many string axions, as generically predicted by string theory.
For example, with $f_\phi = 10^{15}\,\mathrm{GeV}$ and $H_{\rm inf} = 10^{13}\,\mathrm{GeV}$, the existence of around ${\cal O}(100)$ light axions makes it natural that one of them lies near the hilltop due to the stochastic distribution.
Alternatively, the hilltop configuration can be realized through mixing between axions—or phases in the Higgs sector—as studied in Refs.~\cite{Daido:2017wwb,Co:2018mho,Takahashi:2019pqf,Nakagawa:2020eeg,Narita:2023naj,Co:2024bme}.
Another possibility involves matter effects that effectively reverse the potential~\cite{Gonzalez:2022mcx} (see also Ref.~\cite{Khoury:2025txd}).

We also note that in this scenario, the formation time of domain walls can be significantly later than the maximal estimate shown in Fig.~\ref{fig:1}, if the fluctuations are sufficiently small.  
This behavior is illustrated in Fig.~\ref{fig:2}, where we show the evolution of a homogeneous axion field $\phi$ in the left panel and the redshift at the domain wall formation $z_\mathrm{form}$ depending on the initial condition $\phi_\mathrm{ini}/f_\phi$ in the right panel.
Here, we numerically solve the equation of motion  
\beq\laq{eom}
\left((1+z) H \frac{d}{dz}\right)^2\phi - 3H^2(1+z) \frac{d}{dz}\phi = -\partial_\phi V.
\eeq  
The initial field values are taken as $\phi_{\rm ini} = \{10^{-9},\, 1,\, 10^9,\, 10^{13}\}\,\GEV$, arranged from bottom to top in the plot, with $f_\phi = 10^{15}\,\GEV$ and $\sigma_{\rm DW} = (1\,\MEV)^3$.  
These initial values may be considered to be of order $H_{\rm inf}$.  
The time of domain wall formation is identified as the moment when the field reaches $\phi/f_\phi \sim \pi$, and the oscillations begin.

We see that, in order to avoid the domain wall problem for a string axion, the domain walls must form after the recombination era.  
As a result, we do not expect anisotropic birefringence in the CMB.  
However, if the walls form well before reionization, $\phi$ lies in different vacua in different domains at the reionization, and thus photons from that epoch can still experience anisotropic birefringence.  
As a result, CMB photons from the reionization epoch experience anisotropic CB with a rotation angle comparable to the isotropic CB for photons from the recombination epoch.
Since the reionization contribution to the CMB polarization anisotropies is limited, such a signal is more difficult to probe than the anisotropic signal for photons from the recombination epoch~\cite{Takahashi:2020tqv, Kitajima:2022jzz,Gonzalez:2022mcx}.
Current observational constraints on such anisotropic CB remain consistent even with scenarios involving cosmic strings while substantial improvement is expected in future full-sky CMB experiments~\cite{Namikawa:2024sax}.

In addition, if the domain walls formed significantly earlier than the present time, they could have emitted gravitational waves as discussed in Ref.~\cite{Ferreira:2023jbu}. 
They also evaluate the spectrum of the gravitational waves and discuss the CMB constraints on such gravitational waves. 
Such gravitational waves are searched for alongside CMB anisotropies induced by the gravitational potential of domain walls~\cite{Zeldovich:1974uw,Sousa:2015cqa,Lazanu:2015fua,Ramberg:2022irf}.

If the domain walls have only just formed today, they would not have reached the scaling regime.
In that case, many closed domain walls could exist within the current horizon.
This is possible if the wall thickness is much smaller than the horizon size, which can occur when the onset of oscillations is significantly delayed by the anharmonic effect near the hilltop.
Photons from galaxies at relatively low redshift may experience a polarization rotation if they cross domain walls an odd number of times.
The resulting rotation angle is predicted to be $0.42\,c_\gamma\,\mathrm{deg}$, which is twice the isotropic birefringence angle observed in the CMB.

\section{Time-Varying Dark Energy} 

Motivated by the interpretation of the isotropic CB, we consider an axion initially located near the hilltop of its potential, where the anharmonic effect could delay the onset of oscillations.
With such an initial condition, the axion remains near the hilltop for an extended period and starts oscillating at late times.
If the energy density associated with these delayed oscillations is sufficiently large, the axion can contribute to a time-dependent dark energy component.
This enhancement in energy density arises from the anharmonic effect near the potential maximum, which increases the axion abundance compared to the harmonic case.
In particular, for a string axion with a decay constant in the typical range \eq{string}, such an enhancement is crucial—without it, the axion would make only a small contribution to the present dark energy.
Thus, the hilltop initial condition plays an essential role not only in explaining the isotropic CB, but also in realizing a time-dependent dark energy component compatible with the DESI observations.

If domain walls are formed, however, it becomes challenging to simultaneously satisfy the constraint on their tension—imposed by the requirement that their energy density remain below roughly $10^{-5}$ of the present total energy density—and realize a significant deviation of the dark energy equation of state from that of a cosmological constant ($w = -1$).
Although it is in principle possible to suppress the energy stored in the domain walls while enhancing the energy in the bulk oscillations of the axion field, which we will comment on later, we here focus on the implications of a hilltop initial condition for a string axion—suggested by the isotropic CB—for the DESI DR2 observations, assuming that domain walls are not formed.

\begin{figure}[!t]
\begin{center}   \includegraphics[width=140mm]{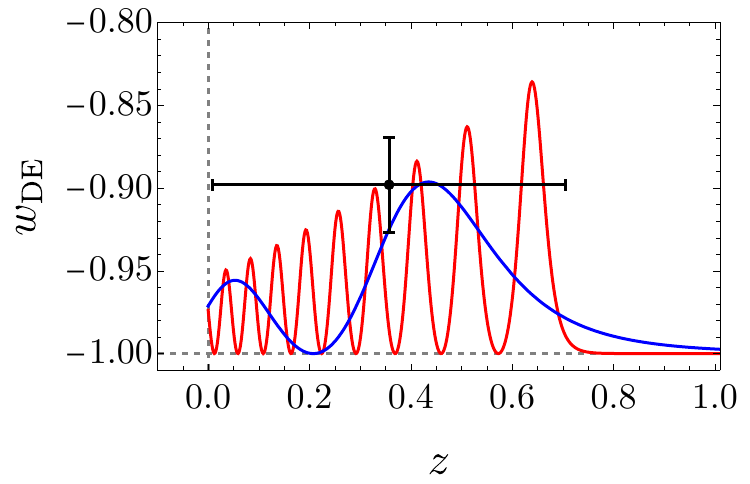}
\end{center}
\caption{%
    Redshift dependence of the dark energy equation of state parameter $w_{\rm DE}(z)$ in the hilltop axion scenario.
    The red (blue) line corresponds to $f_\phi = 10^{16}\,\mathrm{GeV}$, $\phi_{\rm ini} = 3 \,\mathrm{GeV}$, and $m_\phi = 1.13 \times 10^{-31}\,\mathrm{eV}$ ($f_\phi = 6 \times 10^{16}\,\mathrm{GeV}$, $\phi_{\rm ini} = 10^{15}\,\mathrm{GeV}$, and $m_\phi = 1.89 \times 10^{-32}\,\mathrm{eV}$).
    The black point shows the DESI DR2 result from Refs.~\cite{DESI:2025zgx,DESI:2025fii}.
    The horizontal dashed line represents $w_\mathrm{DE} = -1$, corresponding to the cosmological constant, and the vertical dashed line represents the current time, $z = 0$.
}
\label{fig:4}
\end{figure}

To analyze this scenario, one can simulate the system using a homogeneous scalar field by solving \Eq{eom}.
Here we use
\begin{align}
    H=H_0 \sqrt{\Omega_\Lambda+\frac{V+\dot{\phi}^2/2}{3M_{\rm pl}^2H_0^2} + \Omega_m (1+z)^3}
\end{align}
with $M_{\rm pl} = 2.4 \times 10^{18}\,\GEV$ the reduced Planck mass, and $H_0 \approx 68.5\,\rm km/s/{\rm Mpc}$, $\Omega_\Lambda \sim 0.7$, and $\Omega_m \approx 0.3$~\cite{DESI:2025zgx}.
We compute the dark energy equation-of-state parameter as
\begin{equation} 
w_{\rm DE} \equiv \frac{\dot{\phi}^2/2 -V -3 M_{\rm pl}^2 H_0^2 \Omega_\Lambda}{\dot{\phi}^2/2 +V +3 M_{\rm pl}^2 H_0^2 \Omega_\Lambda}.
\end{equation}
In Fig.~\ref{fig:4}, we show the redshift dependence of the dark energy equation of state parameter $w_{\rm DE}(z)$ in the hilltop axion scenario under consideration.
For the red (blue) line, we set $f_\phi = 10^{16}\,\mathrm{GeV}$, $\phi_{\rm ini} = 3 \,\mathrm{GeV}$, and $m_\phi = 1.13 \times 10^{-31}\,\mathrm{eV}$ ($f_\phi = 6 \times 10^{16}\,\mathrm{GeV}$, $\phi_{\rm ini} = 10^{15}\,\mathrm{GeV}$, and $m_\phi = 1.89 \times 10^{-32}\,\mathrm{eV}$), and compare it with the data point from DESI DR2~\cite{DESI:2025zgx,DESI:2025fii}.
While our scenario predicts the oscillation of $w_\mathrm{DE}$ at low redshift, the DESI data can constrain $w_\mathrm{DE}$ only for binned redshifts due to the limited number of tracers for BAO measurements.
We therefore treat the DESI measurements as bin-averaged values of $w_\mathrm{DE}$ and present the qualitative comparison in Fig.~\ref{fig:4}.
Our prediction remains consistent with the $\Lambda$CDM model for $z > 1$, showing no significant deviation in that range.%
\footnote{To obtain a better fit at $z > 1$, modifications to the matter or radiation sector may be required, which are model-dependent. 
We also note that the $z < 0.7$ bin has the smallest error bar and that the deviations from the $\Lambda$CDM for higher redshift bins are less significant ($\lesssim 2 \sigma$) than the $z < 0.7$ bin (see Fig.~12 of Ref.~\cite{DESI:2025zgx}).
Since we do not perform a statistical fit in this paper, the comparison in Fig.~3 should be regarded as qualitative.
}

While we evaluate $w_{\rm DE}$ using a homogeneous field, in reality the axion field develops spatial inhomogeneities after the onset of oscillations due to tachyonic instabilities.
As a result, the local equation-of-state parameter tends to be larger than that of non-relativistic matter.
Nevertheless, since the current observational uncertainties in the DESI measurements are still relatively large, such differences are not expected to be critical at this stage.
A more accurate determination of the effective $w_{\rm DE}$ would require lattice simulations, which are left for future work.

\begin{figure}[!t]
\begin{center}  
   \includegraphics[width=145mm]{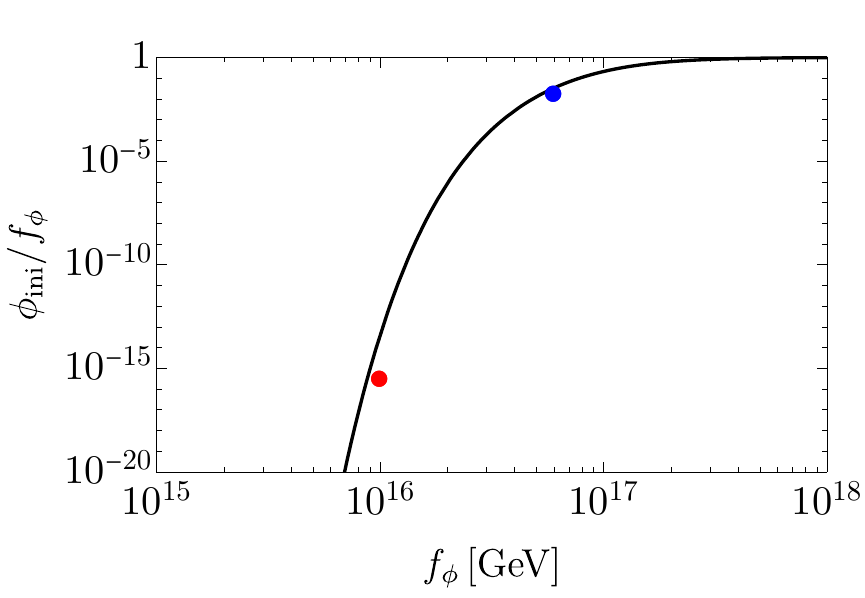}
\end{center}
\caption{%
    Tuning of the initial field value required to achieve $1 + w_{\rm DE} = \mathcal{O}(0.1)$ at redshift $z \sim 0.7$.
    For smaller $f_\phi$, the axion must start increasingly closer to the hilltop, and the required tuning becomes exponentially more severe.
    The red and blue dots correspond to the red and blue lines in Fig.~\ref{fig:4}, respectively.
}
\label{fig:tune}
\end{figure}

Finally, we evaluate the tuning of the initial condition required to explain the DESI result.
To see this, we require that the axion starts to roll down the potential at $z_\mathrm{osc} \simeq 0.7$ and has an energy density of $\mathcal{O}(10)\%$ of the rest of dark energy at that time.
The latter condition can be written as 
\begin{align}
   V(0) \simeq 2 m_\phi^2 f_\phi^2 
    \sim
    0.3 \Omega_\Lambda  M_\mathrm{Pl}^2 H_0^2 
    \ .
\end{align}
At the hilltop, the evolution of the axion can be approximated by 
\begin{align}
    \ddot{\phi} + 3 H \dot{\phi} - m_\phi^2 \phi
    \simeq
    0
    \ .
\end{align}
With $\Omega_\Lambda \sim 0.7$ and $\Omega_m \simeq 0.3$, the dark energy starts to dominate the universe at $z \sim 0.3 < z_\mathrm{osc}$.
Thus, we use the relation during the matter-dominated era, $H = 2/(3t)$, and obtain the solution of
\begin{align}
    \phi
    =
    \phi_\mathrm{ini} \frac{\sinh (m_\phi t)}{m_\phi t} 
    \ ,
\end{align}
which satisfies the initial condition, $\phi(t=0) = \phi_\mathrm{ini}$ and $\dot{\phi}(t=0) = 0$.
Consequently, the former condition can be written as
\begin{align}
    \phi_\mathrm{ini} \frac{\sinh (m_\phi t_\mathrm{osc})}{m_\phi t_\mathrm{osc}} 
    \sim
    f_\phi
    \ ,
\end{align}
where we define $t_\mathrm{osc}$ by $t_\mathrm{osc} \equiv 2/(3H_{\Lambda\mathrm{CDM}}(z_\mathrm{osc}))$ with $H_{\Lambda\mathrm{CDM}}(z_\mathrm{osc}) \equiv H_0 \sqrt{\Omega_\Lambda + \Omega_m (1+z_\mathrm{osc})^3}$.
From these conditions, we obtain the required tuning of the initial condition for fixed values of $f_\phi$ (or $m_\phi$) as shown in Fig.~\ref{fig:tune}.
We also show the parameters from Fig.~\ref{fig:4} by red and blue dots in Fig.~\ref{fig:tune}, which roughly match the analytical estimate.
See, e.g., Refs.~\cite{Gonzalez:2022mcx,Khoury:2025txd,Daido:2017wwb,Co:2018mho,Takahashi:2019pqf,Nakagawa:2020eeg,Narita:2023naj,Co:2024bme} for mechanisms that realize such a hilltop initial condition. For instance, thermal or matter effects could make the potential maximum the minimum temporally in the early universe\cite{Gonzalez:2022mcx,Khoury:2025txd}, and the axion can be stabilized there, thereby realizing the hilltop initial condition.

So far in this section, we have focused on the case in which domain walls do not form, and the axion begins to oscillate coherently around a single minimum.
Before closing this section, let us briefly comment on a possible scenario where domain walls do form, but their energy density is suppressed while retaining a sizable contribution to dark energy from the axion oscillation in the bulk.
To achieve this, the formation of domain walls must be significantly delayed compared to the typical onset time given by Eq.~\eqref{osc}.
In that case, the energy stored in the domain walls is expected to be much smaller than the bulk axion oscillation energy.
If the formation is sufficiently delayed, it may be possible to satisfy the bound required to avoid the domain wall problem (see Eq.~\eqref{tension}), while still inducing a non-negligible deviation in the dark energy equation of state parameter.

\section{Summary}

In this paper, we have investigated the late-time formation of domain walls associated with the string axion as a natural explanation of the isotropic CB, which is motivated by the recent ACT DR6 results~\cite{Louis:2025tst} and the previous analyses of the Planck and WMAP data~\cite{Eskilt:2022cff,Minami:2020odp,Diego-Palazuelos:2022dsq,Eskilt:2022wav,Cosmoglobe:2023pgf}.

Considering the cosmological domain wall problem, domain walls of the string axion should be formed well after the recombination (Fig.~\ref{fig:1}).
Such late formation can be realized if the axion has a hilltop initial condition (Fig.~\ref{fig:2}).
As a result, the isotropic CB angle is robustly predicted as $\beta \approx 0.21 
c_\gamma$\,deg, which agrees well with the recent ACT (and combined) result for $c_\gamma \sim 1$.
While anisotropic CB is suppressed for the CMB photons from the recombination epoch due to the delayed domain wall formation, it is present for photons emitted at a later epoch, e.g., reionization.
In addition, the domain walls themselves can induce the CMB anisotropies and gravitational waves.

Furthermore, we explored the implications of this scenario for time-varying dark energy, motivated by the recent DESI DR2 results~\cite{DESI:2025zgx}, which further support a deviation from a cosmological constant in addition to DESI DR1~\cite{DESI:2024mwx}.
We find that the equation-of-state parameter of dark energy can fit the DESI results better than the $\Lambda$CDM for low redshifts, $z \lesssim1$.

In summary, our scenario highlights the intriguing possibility that late-time onset of string axion oscillations and the subsequent formation of domain walls could simultaneously explain isotropic CB and time-varying dark energy.
To further assess the viability and testability of this scenario, future investigations on the CMB anisotropies, anisotropic CB, and gravitational waves will provide valuable insights, which we leave for future research.

\section*{Acknowledgments}
This work is supported by JSPS Core-to-Core Program (grant number: JPJSCCA20200002) (F.T.), JSPS KAKENHI Grant Numbers 20H01894 (F.T.), 20H05851 (F.T. and W.Y.), 22H01215 (W.Y.), 22K14029 (W.Y.), 23KJ0088 (K.M.), and 24K17039, (K.M.), Graduate Program on Physics for the Universe (J.L.) and JST SPRING Grant Number JPMJPS2114 (J.L.),  Selective Research Fund for Young Researchers from Tokyo Metropolitan University (W.Y.). 
This article is based upon work from COST Action COSMIC WISPers CA21106, supported by COST (European Cooperation in Science and Technology).

\bibliographystyle{apsrev4-1}
\bibliography{ref}
\end{document}